# ITERATIVE LEARNING CONTROL - GONE WILD

S. R. Koscielniak*, TRIUMF, Vancouver, B.C., Canada

*Abstract*

Before AI and neural nets, the excitement was about iterative learning control (ILC): the idea to train robots to perform repetitive tasks, or train a system to reject quasi-periodic disturbances. The excitement waned after the discovery of "learning transients" in systems which satisfy the ILC asymptotic convergence (AC) stability criteria. The transients may be of long duration, persisting long after eigenvalues imply they should have decayed, and span orders of magnitude. They occur both for causal and non-causal learning.

The field recovered with the introduction of tests for "monotonic convergence of the vector norm", but no deep and truly satisfying explanation was offered. Here we explore solutions of the ILC equations that couple the iteration index to the within-trial sample index. This sheds light on the causal learning – for which the AC test gives a repeated eigenvalue.

Moreover, since 2016, this author has demonstrated that a new class of solutions, which are soliton-like, satisfy the recurrence equations of ILC and offer additional insight to long-term behaviour. A soliton is a wave-like object that emerges in a dispersive medium that travels with little or no change of shape at an identifiable speed. This paper is the first public presentation of the soliton solutions, which may occur for both causal (i.e. look back) and noncausal (i.e. look ahead) learning functions that have diagonal band structure for their matrix representation.

## INTRODUCTION

The TRIUMF-Elinac[1,2] is a 30 MeV, 10 mA c.w. capable 1.3 GHz SRF electron linear accelerator. The linacne has a 300 keV electron gun in an SF6-filled tank, a 10 MeV injector cryomodule with one 9-cell cavity, and a 20 MeV accelerator cryomodule (ACM) with two cavities, and a target station; all linked by beam transports. The linac is extensible to 50 MeV by the installation of a second ACM.

The E-linac is nominally a c.w. accelerator, but must operate at much lower power for commissioning and beam development. So e-linac is pulsed with a variety of repetition rate and pulse length, leading to significant transient beam-loading of the SRF cavities. Feed-forward compensation (FFC) of beam-loading transients is essential. Initially, the RF group maintained an argument that the beam would pass before the RF transients grew, and that FFC was not needed; later they conceded the need for FFC.

A different FFC[t] function is needed for every combination of repetition rate and pulse length. In 2013, LLRF group proposed [3] to generate the many FFC[t] time functions by Iterative Learning [4]. Based on numerical simulation, they chose a 4-term noncausal Iterative Learning Control (ILC): **Q**=**I** and **L**=**I**+⅓(↑+↑↑+ ↑↑↑). Here **Q** is a filter and ↑ is a lift applied to the FF vector. But numerical cases do not guarantee stability or convergence, because they cannot span the entire space of initial conditions. So present author, embarked on a complete analysis, culminating in the discovery in 2016 of wave solutions and prediction of unstable learning.

The E-linac LLRF system is a digital system, and well suited to implementing ILC. The RF waveform is demodulated and sampled. The control signals are implemented digitally. All samples can be recorded and held as supervectors.

## ITERATIVE LEARNING CONTROL

System (or plant P) with its own internal (stable) response. Place a ILC wrapper around P and iterate wrapped system from one trial to next. During the trial, the plant is a free system with a driven input. At the end of an individual trial, the input is updated based on results from the ILC wrapper. "Learning" is a matrix-map iteration of input to output vector. Nested within the matrix is the internal response of the plant during a trial, which is different each trial. In time domain the iterations are non-linear. Hopefully the plant and its inputs & outputs and the map all "settle down" so that the wrapped system converges.

ILC occurs in a 2-dimensional space: within a trial, index $k$; and from one iteration to the next, index $j$ or $n$.

*Symbols & Equations*

Vectors (internal index k): **u** = input, **y** = output, **d** = disturbance (repetitive), $\mathbf{y_d}$ = desired output (repetitive), $\mathbf{e_j}=(\mathbf{y_d}-\mathbf{y})_j$ = error. Operators: **P** = "the plant", i.e. the system, **Q** = filter, **L** = learning function.

During the trial: $\mathbf{y_j}=\mathbf{P_u u_j} + \mathbf{P_d d}$. From one trial to the next, "**u**" learns from the previous trial (or trials). $\mathbf{u_{j+1}}=\mathbf{Q}[\mathbf{u_j}+\mathbf{L e_j}]$ where $\mathbf{e_j}=(\mathbf{y_d}-\mathbf{y})_j$. Elimination of **e** leads to iterative maps for **u** and **y**. These maps are function of the internal gains (K) and time constants (τ) of **P**, and the iteration gains (ν) of **L**; or the set {K,τ;ν} for short.

*Converges to what?*

Mapping: operator M such that $x_{n+1}=M:x_n$ where $x_n$ may be scalar or vector. If M=M[n], mapping is non-linear. <u>Mappings have fixed points</u> which may be stable or unstable, and are not necessarily zeros of x. Within the basin of attraction, a map will (eventually) converge on a stable fixed point. Convergence is not a synonym for "stable". "Stability" answers the question: "what happens when x is perturbed from its fixed point?". The perturbations are infinitesimal, the system is locally linear, and the response is a decaying oscillation if "stable".

---

* shane@triumf.ca

Iterations[†] converge on the fixed points (FP):

**[I-Q(I-LP$_u$)]u = QL(y$_d$-P$_d$d),**

**[I+P$_u$(I-Q)$^{-1}$QL ]y = P$_d$d + P$_u$(I-Q)$^{-1}$QLy$_{d'}$**

**[I+P$_u$(I-Q)$^{-1}$QL ]e = y$_d$-P$_d$d.**

In the case of **Q=I**, fixed points simplify: **y=y$_{d'}$ u=P$_u^{-1}$(y$_d$-P$_d$d),** and **e=0,** zero residual error. Iteration about the fixed points is governed by: **ũ$_{j+1}$=Q[I – LP]ũ$_j$** and
**ỹ$_{j+1}$=PQ[I – LP]P$^{-1}$ ỹ$_j$ = [PQP$^{-1}$ – PQL]ỹ$_j$**
The tilde quantities are "differences" relative to the FP.

Matrices of causal operators commute. If the transfer functions **Q,P** are causal, then
**ũ$_{j+1}$= Q[I – LP]ũ$_j$** and **ỹ$_{j+1}$ = Q[I – PL]ỹ$_j$.**

*Properties of ILC maps*

Usually it is stated (or assumed) that ILC maps generate sequences **x$_n$** that exist in a Banach space X; that is **x$_n$** is d-Cauchy convergent and the fixed point **x** is in X. d-Cauchy means: for every positive real number r > 0 there exists some index N such that the distance d(x$_m$,x$_n$)<r whenever m and n are greater than N. This implies ‖**x$_n$** -**x**‖=0 as n → infinity where **x** is in X. The d-Cauchy property implies "eventually convergent", or asymptotically convergent.

*Noncausal Learning Functions*

Sufficiently close to its fixed point, iterants x$_n$ are exponential series (λ$^n$). **ỹ$_{j+1}$= λỹ$_j$** implies the eigenvalue equation **Q[I – PL]ỹ$_j$ = λI ỹ$_j$** The limit of Convergence (of the iterations) means finding conditions {K,τ;ν} such that all eigenvalues (λ) of **A=Q[I – PL]** have modulus < 1. Of course, you may search for values {K,τ;ν} such that |λ|< γ <1. The condition |λ(**A**)|< γ <1 implies permissible {K,τ;ν}. Note: the within-trial motion has been decoupled from the iterations. Coincidentally, the condition for exponential convergence is also the condition for Asymptotic Convergence (AC) of noncausal learning.

Away from the fixed point, other sequences x$_n$ may satisfy the mapping. People have found them experimentally: the so-called learning transients. They persist long after the geometric sequences (λ$^n$) should have decayed. They are consistent with d-Cauchy convergence. They occur both for causal and noncausal learning functions. The "bad" initially diverge to large values before eventually converging. The "worst" initially converge before diverging.

*Causal Learning Functions*

Suppose you look for solutions to the iteration equations in the form **ỹ$_{j+n}$= λ$^n$ ỹ$_j$** where **Q[I – PL]ỹ$_j$ = λI ỹ$_j$**

When **Q, P, L** are causal, **A=Q[I-PL]** takes on the Toeplitz form leading to a single repeated eigenvalue. In such case, iterants do not follow simple geometric series. Solutions, sol[k,n], of the iteration equations may be obtained as the product of exponentials and power series:

$$\text{sol}[k,n] == \lambda^n \sum_{m=1}^{k} n^{k-m}\lambda^{-k+m} a[k,m]\text{sol}[m,0]$$

So, if you were expecting a geometric series, all solutions appear to be learning transients. The within-trial behavior is coupled to the single eigenvalue.

Each element, sol[k], of the solution vector is coupled to preceding elements sol[k-m] & m<k-1. The Asymptotic Convergence condition: As n→∞ each element sol[k,n] converges to zero provided the single eigenvalue |λ|<1. However, the element # k does not start to converge until n>k; with 1≤k≤N where N is matrix size.

The solution procedure is to insert the single known eigensolution sol[1,n] into the second row, and solve for the second eigensolution sol[2,n]; and then continue to the next row and insert all prior solutions.

This form sol[k,n] receives little (or no) attention in the ILC literature. Therefore, we report two simple examples to illustrate the method.

Simple low-dimension example: **Q=L=I** and plant

$$\mathbf{P}=\begin{pmatrix} A & 0 & 0 \\ AB & A & 0 \\ AB^2 & AB & A \end{pmatrix} \text{ and } \mathbf{x}[n]=\begin{pmatrix} x1[n] \\ x2[n] \\ x3[n] \end{pmatrix}$$

The repeated eigenvalue is λ=1-A, and the corresponding eigensolution is $x1[n] \to \lambda^n x1[0]$. The other solutions are $x2[n] \to -ABn\lambda^{-1+n}x1[0] + \lambda^n x2[0]$ and

$$x3[n] \to \frac{1}{2}\lambda^{-2+n}(A^2B^2(-1+n)nx1[0]$$
$$- 2ABn\lambda(Bx1[0] + x2[0])$$
$$+ 2\lambda^2 x3[0])$$

Simple low dimension example: **L=I+↓**

$$\mathbf{L}=\begin{pmatrix} 1 & 0 & 0 & 0 \\ 1 & 1 & 0 & 0 \\ 0 & 1 & 1 & 0 \\ 0 & 0 & 1 & 1 \end{pmatrix} \text{ and } \mathbf{P}=\begin{pmatrix} A & 0 & 0 & 0 \\ AB & A & 0 & 0 \\ AB^2 & AB & A & 0 \\ AB^3 & AB^2 & AB & A \end{pmatrix}$$

The repeated eigenvalue is λ=1-A, and the corresponding eigensolution is $x1[n] \to \lambda^n x1[0]$. Other solutions found after successive back-substitution of previous solutions.

$$x2[n] \to \lambda^{-1+n}(-A(1+B)nx1[0] + \lambda x2[0])$$
$$x3[n] \to \frac{1}{2}\lambda^{-2+n}(A^2(1+B)^2(-1+n)nx1[0]$$
$$- 2A(1+B)n\lambda(Bx1[0] + x2[0])$$
$$+ 2\lambda^2 x3[0])$$
$$x4[n] \to \frac{1}{6}\lambda^{-3+n}(-A^3(1+B)^3 n(2-3n+n^2)x1[0]$$
$$+ 3A^2(1+B)^2(-1+n)\lambda(2B$$
$$+ x2[0])$$
$$- 6A(1+B)n\lambda^2(B^2x1[0] + Bx2[0]$$
$$+ x3[0]) + 6\lambda^3 x4[0])$$

Notice that A=(1-λ). By inspection, if the cross couplings introduced by **L** are not too strong, the leading terms of sol[k,n] are of the form:

$$\text{sol}[k,n] = \sum_{m=1}^{k} \frac{n^{k-m}(1-\lambda)^{-k+m}\lambda^{-k+m+n}b[k,m]}{(k-m)!}$$

where the coefficients b[k,m] are typically powers of a quantity of order unity or smaller. This form with factorial (*k-m*)>>1 in the denominator is strongly suggestive of long-term convergence. Note that b[k,k]=1.

---

[†] If they converge

## Monotonic Convergence of the Norm

Worried about "bad learning behavior", many authors have utilised a more robust form of convergence: leading to more stringent conditions on {K,τ;ν}. Let $S \equiv A^T A$. The inequality $|Au| \leq \sigma |u|$ where σ is maximum $|\lambda|$ of Eigenvalues[S] implies <u>monotonic convergence</u> (MC) of the vector norm $|u|=\text{Sqrt}[u^T \cdot u]$.

The ILC system behaves according to the eigenvalues of **A** with {K,τ;ν} chosen from the condition |Eigenvalues[S(K,τ;ν)]|<1 . For this subset {K,τ;ν}, convergence is stronger; and so the learning transients are reduced or eliminated.

## WHAT OTHER SOLUTIONS OCCUR?

The ILC equations are nonlinear, and 2-dimensional, inviting the question: what other solutions (i) in the subspace between the AC and MC conditions, (ii) inside the space of MC conditions, (iii) away from fixed point, (iv) that couple within-trial behavior to iteration index. To initiate the answer: experiment by direct iteration, see what happens…

### Simple Plant

The "plant" is an RF cavity (time constant $\tau_c = 1/a$) sandwiched by a zero-order hold (ZOH) sampling system (rate $\rho_s = 1/\tau_s$) with a PI (proportional gain Kp & integral gain Ki per second) controller. Let U be the product $U=a.\tau_s = \tau_c \rho_s$. When pole-zero cancellation is used (as here) for the PI controller, then Ki=a.Kp. The plant is wrapped by an iterative learning control (ILC) with gain ν; see Fig.[1]

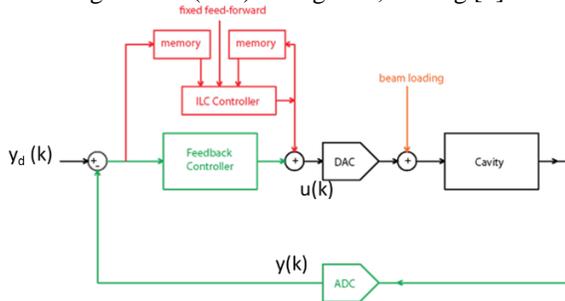

Figure 1: schematic of the E-linac LLRF of RF cavity with closed-loop feedback and a feed forward wrapper and an ILC controller.

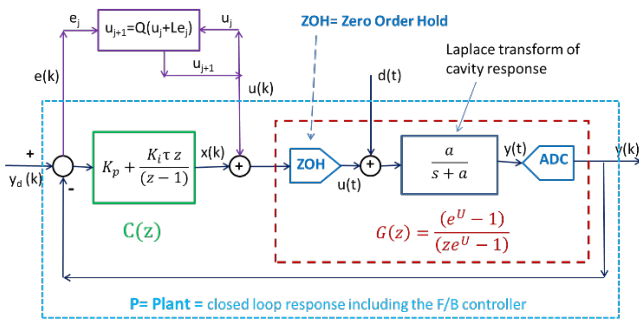

Figure 2: Signal diagram for the plant formed of a RF cavity resonator (red), proportional and integral feedback (green); and its ILC wrapper (purple).

The corresponding signal diagram is shown in Fig.[2]. All the following deals with the single pole problem when Ki=0. Analysis is simplified if we transform from variables U and Kp to A and B, as follows:
$$A = 1 - e^{-U} \text{ and } B = e^{-U}(1 + Kp) - Kp.$$
The ranges of (A,B) are $0 < A < 1$ and $-1 < B < 1$.

### Experiments

The simulation experiments are to choose a simple ILC controller, and then to introduce a disturbance, and then to iterate the equations numerically. Various measure of convergence, such as norm, are extracted from the data. The initial state of the vector **u** or **y** may be either (i) uniform filling of the vector elements with 1's, or (ii) a delta-function at one element and all others zero.

Typically, we witness very many iterations. Iteration of the eigenvalues gives a sense of the naïve expectations of convergence. Parts of the disturbances that survive much longer after the decay of the iterated $\lambda^n$ eigenvalues; are indications of new objects. See for example Figure [3].

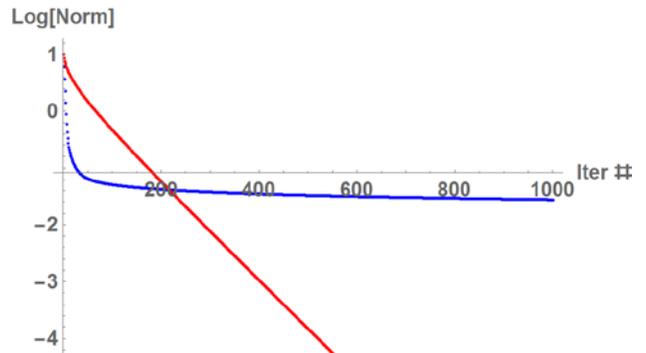

Figure 3: 2-term Look-ahead L=I+↑: Comparison of the vector norm, (blue) from iteration of a uniform-filled initial vector versus power law $\lambda^n$ of the eigenvalues.

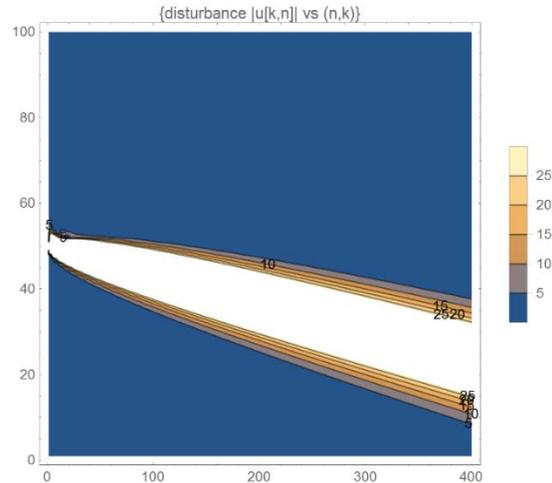

Figure 4: 2-term Look-back L=I+↓: Propagation of a δ-function disturbance in (n,k) space. 1<k<100; 0<n<400.

Figure 4 shows a state vector (internal index *k*) tracked for 400 iterations. It travels in (*n,k*) space at a speed much less than the phase velocity of the ILC algorithm.

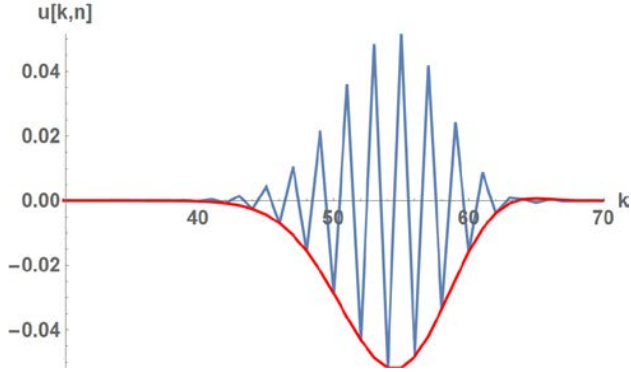

Figure 5: 2-term Look-back L=I+↓: Waveform after 800 iterations: vector element u[k] vs within-trial index *k*.

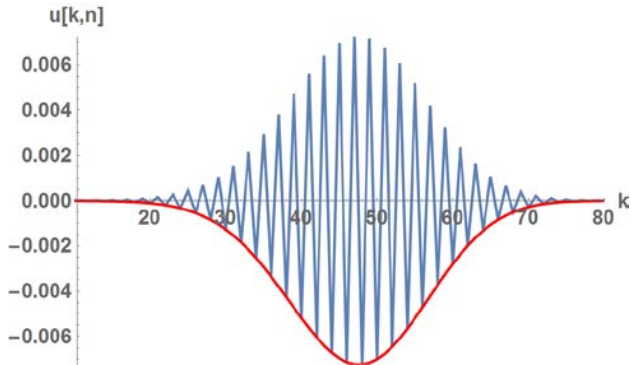

Figure 6: 2-term Look-ahead L=I+↑: Waveform after 800 iterations: vector element u[k] vs within-trial index *k*.

Figures [5,6] are zooms of the state vector over a narrow range of *k* values for causal and noncausal learning, respectively. Both of these residuals, which have persisted long after the decay of iterated eigenvalues, look like waves.

## Wave Packets

Experiments suggest that we are looking for solutions that survive hundreds or more iterations. Compare with the definition of a wave: An object that moves with little or no change of shape at an identifiable speed. Yes, "waves" look like a candidate. Experimentally, it is clear $\mathbf{u_{n+1}} = \mathbf{A^n\, u_1}$ has wave-like solutions for large n -- which become revealed after the geometric sequences have decayed.

Mathematically, the iterants must satisfy a wave equation – something like $\left[\frac{\partial u}{\partial n} + c \frac{\partial u}{\partial k}\right]=0$.
Here *n* = iteration index, *k* = row index of the column vector u, and c = wave velocity. We have to find discrete analogs of the differential operators, and the correct[‡] (soliton-like) waveform.

## Discrete Operators

From the 2-term-learning experiments, we know the wave to be of the form:

$u[k,n] == \mathrm{Cos}\left[\frac{2\pi k}{2}\right] S\left[\frac{k-c\,n}{\sigma}\right]$ where S is the shape, σ is the length scale, and cos(…) is the carrier.

The discrete analog of the differential ∂**u**/∂n is
$$\Delta \mathbf{u}/\Delta \mathbf{n} == (\mathbf{A}-\mathbf{I})\boldsymbol{u} == (\mathbf{Q}(\mathbf{I}-\mathbf{PL})-\mathbf{I})\boldsymbol{u}$$
which becomes Δ**u**/Δ**n** == −**PL** *u* if **Q=I**.
The analog of ∂**u**/∂k is found in element form: Δu/Δk = u[k+1,n]-u[k,n] = Cos[π(k+1)]S[k+1,n] -Cos[πk]S[k,n]. Thus Δu/Δk = -(↑+I)u[k,n] where ↑ is the lift operator acting on a vector arranged vertically with the first element k=1 at the top. For this <u>particular</u> waveform, $\Delta\mathbf{u}/\Delta\mathbf{k} = -(\uparrow + \mathbf{I})\boldsymbol{u}.$

For the differentials, the group velocity is c=(∂u/∂n)/(∂u/∂k). If we have super-vector data **u**, from an experiment[§], then correspondingly

$c = \frac{(A-I)u}{(\uparrow + I)u}$ taken on an element-by-element basis.

To demonstrate wave existence, *ab initio*, we must find the self-consistent waveform **u** and group velocity *c* that satisfies $(\mathbf{A}-\mathbf{I})\boldsymbol{u} == c(\uparrow + \mathbf{I})\boldsymbol{u}.$ A function appears on the left and its derivative on the right, so S must be the exponential form S[X]=Exp[f[X]] where X=(k-k₀)/σ and k₀=c.n is the instantaneous wave centre. The wave neither grows nor decays; therefore f(x)→ - ½ $X^2$.

We insert $u[k,n] == (-1)^k S[-X^2/2]/(2\sigma)$ into the discrete operator wave equation. Comparing each row, k, leads to N equations for speed *c* with only one free parameter, σ. If all equations converge on the same *c* value as σ is varied, then we have a self-consistent solution.

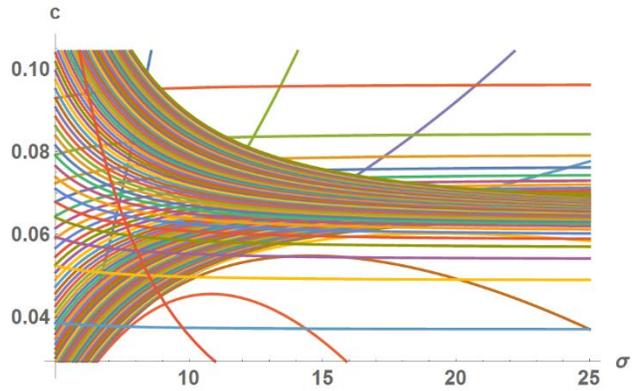

Figure 7: Wave speed, c[k], for each of 1< k<101 equations versus r.m.s width σ for "look-back" **L=I+↓**

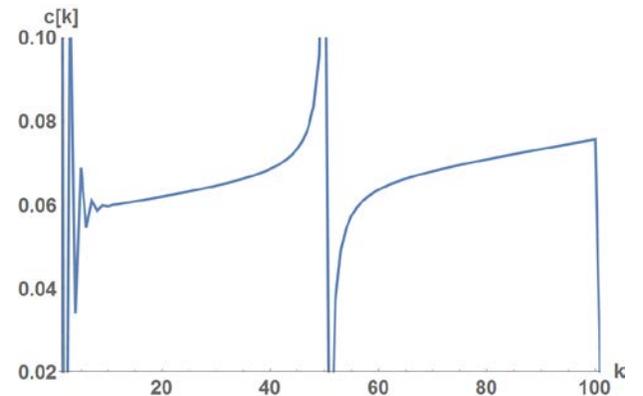

Figure 8: Wave speed, c[k], for each of 1< k<101 equations versus index k for fixed σ=15 for "look-back"**L**

---

[‡] Of course, it is possible there are no such waveforms.

[§] With a 2-term learning function.

Figures [7,8] for causal and Figs. [9,10] for noncausal confirm the existence of wave solutions.

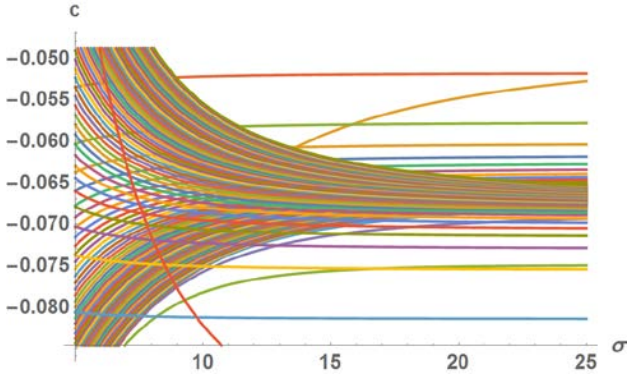

Figure 9: Wave speed, c[k], for each of 1< k<101 equations versus r.m.s width σ for "look-ahead" **L=I+↑**

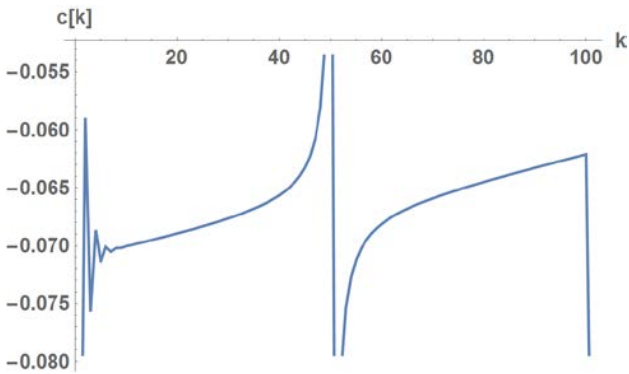

Figure 10: Wave speed, c[k], for each of 1< k<101 equations vs index k for fixed σ=15 for "look-ahead" **L=I+↑**

### Discrete Analogues of ∂u/∂k

The learning function sets the period and form of the carrier, either by advancing or delaying it.

For Identity and 2-term learning, the wave is u[k,n]=Cos[2π(k/2)]S[k,n]. For 3-term learning, wave is u[k,n]={Cos[2π(k/3)]+q1 Cos[2π(k±1)/3]} S[k,n].

For 4-term learning, the wave is u[k,n]= {Cos[2π(k/4)]+q1 Cos[2π(k±1)/4]+q2 Cos[2π(k±2)/4]}S[k,n].

And so on… Take minus sign for look-back; plus for look-ahead. In element form Δu/Δk|$_k$ =u[k+1,n]-u[k,n] .
Substitution of the above, leads to

$$\Delta u/\Delta k|_k = -\frac{1}{2}S[k,n]\left((2+\uparrow -q1 - 2\uparrow q1)\cos\left[\frac{2k\pi}{3}\right] + \sqrt{3}(\uparrow +q1)\sin\left[\frac{2k\pi}{3}\right]\right)$$

for 3-term look-back learning.

In operator form $\frac{\Delta u}{\Delta k} = [r\uparrow -\mathbf{I}]u$ with $r_{k,k} =$

$$= \frac{(1-2q1)\cos\left[\frac{2k\pi}{3}\right] + \sqrt{3}\sin\left[\frac{2k\pi}{3}\right]}{(-2+q1)\cos\left[\frac{2k\pi}{3}\right] - \sqrt{3}q1\sin\left[\frac{2k\pi}{3}\right]}$$

For 4-term look-back L=I+↓+↓²+↓³ learning

$$\Delta u/\Delta k|_k = S[k,n]\left((-1+\uparrow q1 + q2)\cos\left[\frac{k\pi}{2}\right] - (\uparrow +q1 - \uparrow q2)\sin\left[\frac{k\pi}{2}\right]\right)$$

In operator form $\frac{\Delta u}{\Delta k} = [r\uparrow -\mathbf{I}]u$ with

$$r_{k,k} = -\frac{q1\cos\left[\frac{k\pi}{2}\right] + (-1+q2)\sin\left[\frac{k\pi}{2}\right]}{(-1+q2)\cos\left[\frac{k\pi}{2}\right] - q1\sin\left[\frac{k\pi}{2}\right]}$$

## CONCLUSIONS

### Well Known

If convergent, ILC maps iterate to their fixed points (FP). If **Q=I**, then FP=**0**. If **Q≠I**, then FP≠**0** and residual error **e≠0.**

Near the fixed point, noncausal learning has eigen-solutions such that **A$^n$e$_k$=λ$^n_k$e$_k$**. Gains {K,τ;ν} chosen from the MC condition |λ(A$^T$A)|<1 influence λ$_k$ .
Stability conditions for causal and noncausal learning are both subject to P alone is stable for {K,τ;0}.

### Less Well Known

Causal learning has eigen-solutions **A$^n$e$_k$=λ$^n$e$_k$(n)** with **e$_k$**(n)= $\sum_{m=1}^{k} n^{k-m}\lambda^{-k+m}a[k,m]\text{sol}[m,0]$
Gains {K,τ;ν} chosen from the MC condition |λ(A$^T$A)|<1 influence λ & a[k,m]. Eigensolution #k starts to converge for n>k.

### Waves Found In 2016

When **Q=I** and **L** is the diagonal-band style of learning function **L**=ν$_0$**I**+Σ$_q$ν$_q$↑$^q$+Σ$_p$ν$_p$↓$^p$ , there may also be soliton-type wave solutions S(*k-c.n*) existing inside the MC domain. The waves have carrier close to the Nyquist frequency ρ$_s$/2. The waves obey the equation **(A-I)u** = c(**Δ/Δk**)**u** where the wave shape S and speed c must be found self-consistently. Whereas the phase velocity of propagation of the ILC algorithms is c=k/n=1, the group velocity of the wave |c|= |Δ**k**/Δ**u**| <<1.

The waves may be suppressed by a low pass filter **Q**, but at the cost of introducing residual errors, and changing the eigensolutions.

This is a cautionary tale. ILC has 2-dimensions *n* and *k*; focusing only on the ILC wrapper and iteration dynamics *(n)* while ignoring the internal dynamics *(k)* is perilous.